\documentclass[12pt]{iopart}

\begin{document}

\title[Author guidelines for IOP journals in  \LaTeXe]{Impurity Screening and Surface Acoustic Wave Absorption in a Dipolar Exciton Condensate at Finite Temperatures}

\author{V.M. Kovalev$^{1,2}$ and A.V. Chaplik$^{1,3}$}

\address{$^1$Institute of Semiconductor Physics, Siberian Branch of the Russian Academy of Sciences, Novosibirsk, 630090, Russia\\
$^2$Novosibirsk State Technical University, Novosibirsk, 630095, Russia\\
$^3$Novosibirsk State University, Novosibirsk, 630090, Russia}
\ead{vadimkovalev@isp.nsc.ru}
\begin{abstract}
We describe the behavior of a repulsively interacting Bose-Einstein condensate of indirect dipolar exciton gas in a double quantum well (QW) system under external static or dynamic electric fields at finite temperatures. Electrostatic perturbation is considered to be created by an impurity atom or shot-range defect of QW fluctuation. The screening of this defect potential by an exciton condensate is studied. We find asymptotic spatial dependence of the screened potential and analyse its dependence on the temperature and exciton concentration. It is shown that the asymptotic of the screened potential has a steep power law dependence in contrast to the well known results of electron gas. This peculiarity reflects the bosonic nature of the exciton condensate.

The behavior of exciton condensate under external alternative field created by a surface acoustic wave (SAW) is examined in detail. We focus our attention on the dependence of SAW absorption coefficient on temperature and exciton concentration. We found that at zero temperatures Landau damping does not contribute to the SAW absorption, but the Belyaev mechanism produces unusual behavior of SAW absorption coefficient on exciton concentration: if the exciton concentration exceeds some critical value, the SAW absorption vanishes. At finite temperatures Landau damping comes into action and results in washing out the sharp absorption behavior. Such unusual SAW absorption properties can be used for experimental evidence of the exciton condensation. This method is also applicable to the experimental testing of both dark and bright exciton condensates, that is impossible to do with the optical luminescence technique.
\end{abstract}

\maketitle

\section{Introduction}
The two-dimensional indirect excitons (or exciton polaritons) in double and wide single QWs are one of the most prominent systems under study in condensed matter physics. It attracts wide attention because of its unique physical properties. The most important physical phenomenon widely studied both experimentally and theoretically is exciton Bose-Einstein condensation (BEC)~\cite{ButovReview},\cite{GorTimReview}. Indirect excitons, being a bound state of the electron and hole localized in spatially separated QWs, possesses a long life-time comparison with usual excitons. Long life-time of indirect excitons at sufficiently low temperatures allows one to accumulate a large amount of excitons in the lowest quantum state, the BEC phenomenon. Being located in a solid state structure, BEC of dipolar excitons undergoes the influence of imperfections. The effect of disorder on the BEC properties was theoretically studied in literature (see, for example~\cite{CaulMuller} and references therein). In the present work we examine the opposite case - the effect of exciton BEC on the spatial dependence of impurity potential in the QW, e.g. the screening phenomenon. On the one hand, the problem of the local perturbation potential screening by the BEC of neutral particles is a general problem of condensed matter physics. It will be shown below that the screening phenomenon, in this situation, demonstrates sharp distinctions from the known results for electron gas. On the other hand, its solution is of large importance for the explanation of existing experiments~\cite{Butov2}. It should be also noted that the problem of the impurity screening by excitonic gas was studied in literature in the absence of excitonic BEC \cite{KovalevChaplik1}, \cite{Ivanov}.

Experimental evidence of the existence of exciton BEC is mainly based on the optical arguments. We show in the present work that absorption of SAW experimental technique widely used in earlier studies of two-dimensional electron gas may provide with an alternative method to study the dipolar excitons BEC. Moreover, using the SAW technique allows one to detect both dark and bright exciton condensate on the equal footing, which is impossible to do with optical methods,-- in particular by means of the luminescence technique.

In our present paper we suggest the hybrid LiNbO$_3$/DQW structure for studying the exciton BEC by means of SAW. We calculate below the SAW absorption coefficient due to its interaction with condensate and noncondensate exciton particles at zero and finite temperatures. We show that SAW absorption coefficient includes both Belyaev and Landau damping precesses. Landau damping of SAW phonons is absent at zero temperatures, and both Belyaev mechanism and wave-transformation mechanism (direct conversion of SAW phonon to the Bogolubov sound phonon) occur. Microscopic calculation shows that the Belyaev damping process results in an unusual behavior of SAW absorption vs concentration of the exciton condensate, e.g. there is a critical concentration value above which SAW absorption vanishes. The Landau damping of SAW phonons appears at finite temperatures, but it is exponentially small at low temperatures (the exact criterion will be given below). We assume that, even at small temperatures, the presence of exciton BEC can be experimentally tested by measuring SAW absorption.

\section{Dipolar exciton BEC response to the external time-dependent field}
Time-dependent BEC density fluctuations (collective modes) are well studied in literature~\cite{Giorgini1},\cite{Giorgini2}. Depending on the relation between frequency $\omega$ and typical collision time $\tau$, one can distinguish two limiting regimes. Regime $\omega\tau >>1$ is referred to as collisionless  regime, which is described either by Popov effective field theory approach or Hartree-Fock-Bogoliubov (HFB) approximation~\cite{book}. In the opposite case $\omega\tau <<1$ the dynamic behavior of the condensate is described by the theory of two-fluid hydrodynamics. We focus our attention here on collisionless case $\omega\tau >>1$ and apply the time-dependent mean-field approach used S. Giorgini~\cite{Giorgini1} to study the collective excitations damping in the dilute Bose gas.

The starting point is a Hamiltonian describing the Bose particles interacting via point-like potential $W(\textbf{r}-\textbf{r}')=g\delta(\textbf{r}-\textbf{r}')$
\begin{eqnarray}\label{1}
\hspace{-1cm}H=\int d\textbf{r}\Psi^{+}(\textbf{r},t)\left(\frac{\textbf{p}^2}{2M}-\mu+U(\textbf{r},t)\right)\Psi(\textbf{r},t)+g\int d\textbf{r}\left[\Psi^{+}(\textbf{r},t)\Psi(\textbf{r},t)\right]^2.
\end{eqnarray}
The specific form of constants $g$ and $M$ will be given below for the dipolar exciton system. Here we only derive the general formulae for the BEC response to be used later on. The Bose field $\Psi(\textbf{r},t)$ in eq.(\ref{1}) can be decomposed into condensate $\varphi(\textbf{r},t)$ and noncondensate $\psi(\textbf{r},t)$ contributions
\begin{eqnarray}\label{2}
\Psi(\textbf{r},t)=\varphi(\textbf{r},t)+\psi(\textbf{r},t).
\end{eqnarray}
To derive equations of motion for $\varphi(\textbf{r},t)$ and $\psi(\textbf{r},t)$ fields, we use the mean-field scheme described in detail in~\cite{Giorgini1}. Leaving out all calculations we arrive to the dynamic equations for the condensate field
\begin{eqnarray}\label{3}
\hspace{-2.5cm}\left(
  \begin{array}{cc}
    i\partial_t-\frac{p^2}{2M}+\mu-U-g[|\varphi_x|^2+2n_x] & -gm_x \\
    -gm_x^{\ast} &  -i\partial_t-\frac{p^2}{2M}+\mu-U-g[|\varphi_x|^2+2n_x] \\
  \end{array}
\right)\left(
         \begin{array}{c}
           \varphi_x \\
           \varphi_x^{\ast} \\
         \end{array}
       \right)=0~~~~~~
\end{eqnarray}
where $n_x=\langle\psi^{\ast}(x)\psi(x)\rangle$ and $m_x=\langle\psi(x)\psi(x)\rangle$ are normal and anomalous noncondensate particle densities, respectively. Here, short-hand notation $x=(\textbf{r},t)$ is used. The noncondensate particles dynamics is described by the following matrix equation
\begin{eqnarray}\label{4}
\hspace{-2.5cm}\left(
  \begin{array}{cc}
    i\partial_t-\frac{p^2}{2M}+\mu-U-2g[|\varphi_x|^2+n_x] & -g[\varphi_x^2+m_x] \\
    -g[\varphi_x^{\ast 2}+m_x^{\ast}] &  -i\partial_t-\frac{p^2}{2M}+\mu-U-2g[|\varphi_x|^2+n_x] \\
  \end{array}
\right)\hat{G}=\hat{1},
\end{eqnarray}
where the relation between densities $n_x$
and $m_x$ with Green function $\hat{G}$ is
\begin{eqnarray}\label{5}
\left(
  \begin{array}{cc}
    n_x & m_x \\
    m^{\ast}_x & n_x \\
  \end{array}
\right)=\left(
  \begin{array}{cc}
    iG(x) & iF(x) \\
    iF^{+}(x) & i\tilde{G}(x) \\
  \end{array}
\right).
\end{eqnarray}
Here, Green functions are taken at coincident space and time positions. To derive the BEC response to the external potential $U$, we treat $U$ as a small perturbation. Small external field $U$ produces a small-amplitude deviation of field $\varphi_x$ and noncondensate particle densities $n_x,m_x$ from their stationary uniform values
\begin{eqnarray}\label{6}
\varphi_x=\sqrt{n_c}+\delta\varphi_x,\,\,\,|\varphi_x|^2=n_c+\delta n_c;\\
\nonumber
n_x=n_0+\delta n_x,\,\,\,\,\,\,m_x=m_0+\delta m_x,
\end{eqnarray}
where $\delta n_c=\sqrt{n_c}(\delta\varphi_x^{\ast}+\delta\varphi_x).$
Let us first consider the dynamics of field $\varphi_x$. In all equations below we will exploit the so-called HFB-Popov approximation $m_0=0$~\cite{book}. Making a decomposition (\ref{6}) in eq.(\ref{3}), we find the equation of motion for the $\varphi_0=\sqrt{n_c}$ part, which gives the relation between chemical potential $\mu$ and equilibrium densities $\mu=gn_c+2n_0$. The time-dependent equation for $\delta\varphi_x$ is obtained by linearizing eq.(\ref{3})
\begin{eqnarray}\label{7}
\hspace{-2cm}\left(
  \begin{array}{cc}
    i\partial_t-\frac{\textbf{p}^2}{2M}+\mu-g[n_c+2n_0] & 0 \\
    0 &  -i\partial_t-\frac{\textbf{p}^2}{2M}+\mu-g[n_c+2n_0] \\
  \end{array}
\right)\left(
         \begin{array}{c}
           \delta\varphi_x \\
           \delta\varphi_x^{\ast} \\
         \end{array}
       \right)=\\\nonumber
       =\left(
                  \begin{array}{cc}
                    U+g[\delta n_c+2\delta n_x] & g\delta m_x \\
                    g\delta m_x & U+g[\delta n_c+2\delta n_x]\\
                  \end{array}
                \right)\left(
                         \begin{array}{c}
                           \varphi_0 \\
                           \varphi_0 \\
                         \end{array}
                       \right).
\end{eqnarray}
The right-hand side of this equation describes the condensate particles with both external field $U$ and density fluctuations of the noncondensate particles. The last interaction is responsible for damping of Bogoliubov excitations~\cite{Giorgini1}. To simplify our calculations, we disregarded this condensate-noncondensate interaction terms for condensate particles, i.e. the lifetime of Bogoliubov quasiparticles is assumed to be infinite. The equation of motion for $\delta\varphi_x$ in a simplified form can be written as
\begin{eqnarray}\label{8}
\hspace{-2cm}\left(
  \begin{array}{cc}
    i\partial_t-\frac{\textbf{p}^2}{2M}+\mu-g[n_c+2n_0] & 0 \\
    0 &  -i\partial_t-\frac{\textbf{p}^2}{2M}+\mu-g[n_c+2n_0] \\
  \end{array}
\right)\left(
         \begin{array}{c}
           \delta\varphi_x \\
           \delta\varphi_x^{\ast} \\
         \end{array}
       \right)=\\\nonumber
       =\left(
                  \begin{array}{cc}
                    U+g\delta n_c & 0 \\
                    0 & U+g\delta n_c\\
                  \end{array}
                \right)\left(
                         \begin{array}{c}
                           \varphi_0 \\
                           \varphi_0 \\
                         \end{array}
                       \right).
\end{eqnarray}
Let us introduce the polarization operator $P^c_{k\omega}$ of condensate particles by relation $\delta n_c(k,\omega)=P^c_{k\omega}U_{k\omega}$, where $U(\textbf{r},t)=U_{k\omega}e^{i\textbf{kr}-i\omega t}$. $P^c_{k\omega}$ can be easily found with the help of eq.(\ref{8})
\begin{eqnarray}\label{9}
P^c_{k\omega}=n_c\frac{k^2/M}{(\omega+i\delta)^2-\epsilon_k^2},\,\,\,\epsilon_k^2=\frac{k^2}{2M}\left(\frac{k^2}{2M}+2gn_c\right).
\end{eqnarray}
Here, $\epsilon_k$ is a Bogoliubov quasiparticle energy. To find the noncondensate particle density response, the linearization procedure has to be applied to the eq.(\ref{4}) resulting in Dyson equation $\hat{G}_0^{-1}\hat{G}=\hat{1}+\hat{X}\hat{G}$, where
\begin{eqnarray}\label{10}
\hat{G}_0^{-1}=\left(
                 \begin{array}{cc}
                   i\partial_t-\frac{p^2}{2M}+\mu-2g[n_c+n_0] & -gn_c \\
                   -gn_c & -i\partial_t-\frac{p^2}{2M}+\mu-2g[n_c+n_0]  \\
                 \end{array}
               \right)\\\nonumber
\hat{X}=\left(
          \begin{array}{cc}
            U+2g[\delta n_c+\delta n] & g[\delta n_c+\delta m] \\
            g[\delta n_c+\delta m] & U+2g[\delta n_c+\delta n] \\
          \end{array}
        \right)=\\\nonumber
=\left(
  \begin{array}{cc}
    U+2g\delta n_c & g\delta n_c \\
    g\delta n_c & U+2g\delta n_c \\
  \end{array}
\right)+\left(
          \begin{array}{cc}
            2g\delta n & g\delta m \\
            g\delta m & 2g\delta n \\
          \end{array}
        \right).
\end{eqnarray}
Herein, the first matrix in $\hat{X}$ can be considered as an external field having two contributions $U$ and $g\delta n_c$, the induced field of the condensate particle density fluctuations. In the linear response regime, we have to keep only first order terms in Dyson equation, i.e. $\hat{G}\approx \hat{G}_0+\hat{G}_0\hat{X}\hat{G}_0$. $\hat{X}$ matrix includes the noncondensate density fluctuations which have to be found self-consistently. Using the connection between $\delta n,\delta m$ and Green function $\hat{G}$ in the form
\begin{eqnarray}\label{11}
\left(
  \begin{array}{cc}
    \delta n & \delta m \\
    \delta m & \delta n \\
  \end{array}
\right)=i(\hat{G}-\hat{G}_0),
\end{eqnarray}
we arrive to the self-consistent equation to determine noncondensate fluctuation densities
\begin{eqnarray}\label{12}
\left(
  \begin{array}{cc}
    \delta n & \delta m \\
    \delta m & \delta n \\
  \end{array}
\right)=i\hat{G}_0\left(
  \begin{array}{cc}
    U+2g\delta n_c & g\delta n_c \\
    g\delta n_c & U+2g\delta n_c \\
  \end{array}
\right)\hat{G}_0+\\\nonumber+i\hat{G}_0\left(
          \begin{array}{cc}
            2g\delta n & g\delta m \\
            g\delta m & 2g\delta n \\
          \end{array}
        \right)\hat{G}_0,
\end{eqnarray}
where Fourier-transformed $\hat{G}_0$-functions are given by
\begin{eqnarray}\label{13}
\hat{G}_0=\left(
            \begin{array}{cc}
              G_0 & F_0 \\
              F_0 & \tilde{G}_0 \\
            \end{array}
          \right)=\left(
                    \begin{array}{cc}
                      \frac{\omega+k^2/2M+gn_c}{\omega^2-\epsilon_k^2+i\delta} & \frac{-gn_c}{\omega^2-\epsilon_k^2+i\delta} \\
                     \frac{-gn_c}{\omega^2-\epsilon_k^2+i\delta} & \frac{-\omega+k^2/2M+gn_c}{\omega^2-\epsilon_k^2+i\delta} \\
                    \end{array}
                  \right).
\end{eqnarray}
In the small momentum approximation, $\epsilon_k\approx sk$, where $s=\sqrt{gn_c/M}$ is a speed of the Bogoliubov quasiparticles. Such approximation is held under condition $k^2/2M<<2gn_c$ and $\omega\sim sk<<2gn_c$. It permits us to keep only the $gn_c$ term in a matrix Green function eq.(\ref{13}) in the numerators
\begin{eqnarray}\label{14}
\hat{G}_0=\frac{gn_c}{\omega^2-s^2k^2+i\delta}\left(
                                                \begin{array}{cc}
                                                  1 & -1 \\
                                                  -1 & 1 \\
                                                \end{array}
                                              \right).
\end{eqnarray}
This representation for Green functions and relation $\delta n_c=P^cU$ allow us to find the solution of eq.(\ref{12}). After simple, but lengthy algebra, we find the noncondensate particle density response to external field $U_{k\omega}$
\begin{eqnarray}\label{15}
\delta n(k,\omega)=\frac{P^n_{k\omega}[1+gP^c_{k\omega}]}{1-3gP^n_{k\omega}}U_{k\omega},
\end{eqnarray}
where
\begin{eqnarray}\label{16}
P^n_{k\omega}=2i\sum_{\textbf{q},\Omega}F(\textbf{q},\Omega)F(\textbf{q}+\textbf{k},\Omega+\omega)
\end{eqnarray}
is a polarization operator of noncondensate particles. The above-used calculation procedure is the RPA approach applied to account for noncondensate particle interaction contribution to the response equation (\ref{15}). We can see that the total density response $\delta N=\delta n_c+\delta n$ given by eq.(\ref{9}) and eq.(\ref{15}) is determined by the condensate $P^c_{k\omega}$ and noncondensate $P^n_{k\omega}$ polarization operators. $P^n_{k\omega}$ contains the temperature dependence of the BEC density response.

Now let us calculate $P^n_{k\omega}$ at finite temperatures. An exact formula for $P^n_{k\omega}$ in the framework of Matsubara technique is
\begin{eqnarray}\label{17}
P^n_{k\omega_n}=-2T\sum_{\textbf{q},\Omega_m}F(\textbf{q},i\Omega_m)F(\textbf{q}+\textbf{k},i\Omega_m+i\omega_n),
\end{eqnarray}
where
\begin{eqnarray}\label{18}
F(\textbf{q},i\omega_l)=-\frac{gn_c}{2sq}\left(\frac{1}{i\omega_l-sq}-\frac{1}{i\omega_l+sq}\right).
\end{eqnarray}
The summation in eq.(\ref{17}) can be easily done by a standard procedure~\cite{FetterBook}. The result yields
\begin{eqnarray}\label{19}
\hspace{-2cm}P^n_{k,i\omega_n}=-\frac{g^2n_c^2}{2}\int \frac{d\textbf{p}}{(2\pi)^2}\frac{(N_{\bf{p+k}}-N_{\bf{p}})}{\epsilon_{\bf{p+k}}\epsilon_{\bf{p}}}
\left[\frac{1}{i\omega_n-\epsilon_{\bf{p+k}}+\epsilon_{\bf{p}}}-\frac{1}{i\omega_n+\epsilon_{\bf{p+k}}-\epsilon_{\bf{p}}}\right]-\\\nonumber
-\frac{g^2n_c^2}{2}\int \frac{d\textbf{p}}{(2\pi)^2}\frac{(1+N_{\bf{p+k}}+N_{\bf{p}})}{\epsilon_{\bf{p+k}}\epsilon_{\bf{p}}}
\left[\frac{1}{i\omega_n+\epsilon_{\bf{p+k}}+\epsilon_{\bf{p}}}-\frac{1}{i\omega_n-\epsilon_{\bf{p+k}}-\epsilon_{\bf{p}}}\right];
\end{eqnarray}
here and below, we assume that $\epsilon_{\bf{p}}=sp$, and $N_{\bf{p}}=[\exp(\epsilon_{\bf{p}}/T)-1]^{-1}$ is the Bose distribution function. The exciton polarization operator can be separated into temperature-independent and temperature-dependent parts, respectively, $P^n_{k,i\omega_n}=P^{n(0)}_{k,i\omega_n}+P^{n(1)}_{k,i\omega_n}$.
\begin{eqnarray}\label{20}
P^{n(0)}_{k,i\omega_n}=-\frac{g^2n_c^2}{2(2\pi)^2}\int \frac{d\textbf{p}}{\epsilon_{\bf{p+k}}\epsilon_{\bf{p}}}
\left[\frac{1}{i\omega_n+\epsilon_{\bf{p+k}}+\epsilon_{\bf{p}}}-\frac{1}{i\omega_n-\epsilon_{\bf{p+k}}-\epsilon_{\bf{p}}}\right],\\\nonumber
P^{n(1)}_{k,i\omega_n}=-\frac{g^2n_c^2}{2(2\pi)^2}\int \frac{d\textbf{p}N_{\bf{p}}}{\epsilon_{\bf{p+k}}\epsilon_{\bf{p}}}\times\\\nonumber
\times\left[\frac{1}{i\omega_n+\epsilon_{\bf{p+k}}+\epsilon_{\bf{p}}}
-\frac{1}{i\omega_n-\epsilon_{\bf{p+k}}-\epsilon_{\bf{p}}}
-\frac{1}{i\omega_n-\epsilon_{\bf{p+k}}+\epsilon_{\bf{p}}}
+\frac{1}{i\omega_n+\epsilon_{\bf{p+k}}-\epsilon_{\bf{p}}}\right]-\\\nonumber
-\frac{g^2n_c^2}{2(2\pi)^2}\int \frac{d\textbf{p}N_{\bf{p+k}}}{\epsilon_{\bf{p+k}}\epsilon_{\bf{p}}}\times\\\nonumber
\times\left[\frac{1}{i\omega_n+\epsilon_{\bf{p+k}}+\epsilon_{\bf{p}}}
-\frac{1}{i\omega_n-\epsilon_{\bf{p+k}}-\epsilon_{\bf{p}}}
+\frac{1}{i\omega_n-\epsilon_{\bf{p+k}}+\epsilon_{\bf{p}}}
-\frac{1}{i\omega_n+\epsilon_{\bf{p+k}}-\epsilon_{\bf{p}}}\right].
\end{eqnarray}
To proceed further, let us change the variable $\textbf{p+k}\rightarrow\textbf{p}$ in the second brackets in $P^{n(1)}$ and introduce $p_1=|\textbf{p+k}|$. After simple manipulations we get
\begin{eqnarray}\label{21}
\hspace{-2cm}P^{n(0)}_{k,i\omega_n}=-\frac{4g^2n_c^2}{(2\pi)^2}\int_0^{\infty}\frac{pdp}{\epsilon_p}\int_{|p-k|}^{p+k}\frac{p_1dp_1}{\epsilon_{p_1}}
\frac{1}{\sqrt{[(p+k)^2-p_1^2][p_1^2-(p-k)^2]}}\times\\\nonumber
\times\left[\frac{1}{i\omega_n+\epsilon_p+\epsilon_{p_1}}-\frac{1}{i\omega_n-\epsilon_p-\epsilon_{p_1}}\right],\\\nonumber
\hspace{-2cm}P^{n(1)}_{k,i\omega_n}=-\frac{4g^2n_c^2}{(2\pi)^2}\int_0^{\infty}\frac{pdpN_p}{\epsilon_p}\int_{|p-k|}^{p+k}\frac{p_1dp_1}{\epsilon_{p_1}}
\frac{1}{\sqrt{[(p+k)^2-p_1^2][p_1^2-(p-k)^2]}}\times\\\nonumber
\times\left[\frac{1}{i\omega_n+\epsilon_p+\epsilon_{p_1}}-\frac{1}{i\omega_n-\epsilon_p-\epsilon_{p_1}}
-\frac{1}{i\omega_n+\epsilon_p-\epsilon_{p_1}}+\frac{1}{i\omega_n-\epsilon_p+\epsilon_{p_1}}\right].
\end{eqnarray}
here, $\epsilon_p+\epsilon_{p_1}=s(p+p_1)$ and $\epsilon_p-\epsilon_{p_1}=s(p-p_1)$. To calculate the integrals, we introduce new variables $p+p_1=x,\,\,p_1-p=y$. Taking an analytical continuation $i\omega_n\rightarrow i\eta sign[\omega]$ we have
\begin{eqnarray}\label{22}
\hspace{-2cm}P^{n(0)}_{k,\omega}=-\frac{g^2n_c^2\pi}{(2\pi s)^2}\int_k^{\infty}\frac{dx}{\sqrt{x^2-k^2}}\left[\frac{1}{\omega+sx+i\eta sign[\omega]}-\frac{1}{\omega-sx+i\eta sign[\omega]}\right],\\\nonumber
\hspace{-2cm}P^{n(1)}_{k,\omega}=-\frac{2g^2n_c^2}{(2\pi s)^2}\int_k^{\infty}\frac{dx}{\sqrt{x^2-k^2}}\int_{-k}^k\frac{dy}{\sqrt{k^2-y^2}}\frac{1}{e^{s(x-y)/2T}-1}\times\\\nonumber
\hspace{-2cm}\times\left[\frac{1}{\omega+sx+i\eta sign[\omega]}-\frac{1}{\omega-sx+i\eta sign[\omega]}+\frac{1}{\omega+sy+i\eta sign[\omega]}-\frac{1}{\omega-sy+i\eta sign[\omega]}\right].
\end{eqnarray}
Now $Im\,P^{n(0)}_{k,\omega}$ can be calculated via equality $(a+i\eta)^{-1}=a^{-1}-i\pi\delta(a)$. The integral for $Re\,P^{n(0)}_{k,\omega}$ can be manipulated with $x=k\cosh(z)$. The result yields
\begin{eqnarray}\label{23}
P^{n(0)}_{k,\omega}=-\frac{g^2n_c^2}{4s^2}
\left[\frac{\theta(s^2k^2-\omega^2)}{\sqrt{s^2k^2-\omega^2}}+i\frac{\theta(\omega^2-s^2k^2)}{\sqrt{\omega^2-s^2k^2}}\right].
\end{eqnarray}
We leave the $P^{n(1)}_{k,\omega}$ in the integral form of eq.(\ref{22}) because it cannot be found analytically. We use this integral representation to calculate the limiting cases that will be considered in the next sections.

\section{Screening of static potential}
We consider the double quantum well (DQW) structure depicted in Fig. 1. The electron and hole are located in different QWs interacting via Coulomb potential forming the exciton with the dipole moment $\textbf{p}$ directed to the normal of the DQW plane. We will assume the simple exciton model. It will be considered as a rigid dipole particle with a dipole moment along direction $y$ only, $\textbf{p}=(0,-ed,0)$. Here $e$ is an electron charge and $d$ is a distance between QWs. Such a model ignores the internal exciton motion of the particles and its motion in axis $y$ also. This model is good enough to describe the system under study, while the internal degrees of freedom are not excited. We assume that both time-dependent electric field of acoustic wave and especially the static field of the impurity cannot excite them. Nevertheless, the dipoles, as a whole, are free to move in the $(x,z)$ plane. In this section we consider the problem of electrostatic screening and acoustic wave absorption will be examined in the following section.

Static impurity potential $U(\bf{r})$ produces the static exciton density variation which, in turn, gives rise to induced potential $W_{ind}(\bf{r})$. The total field is given by $W(\textbf{r})=U(\textbf{r})+W_{ind}(\textbf{r})$. The Fourier-transform of field $W(k)$ reads
\begin{eqnarray}\label{3.1}
W(k)=U(k)+g[\delta n_c(k,0)+\delta n(k,0)].
\end{eqnarray}
We apply here the linear response theory, where $W_{ind}\sim \delta N$.
Now we need to specify the form of exciton-exciton interaction $g$.
The Fourier-transform of exciton-exciton interaction $g(\textbf{r}-\textbf{r}')$ is given by
\begin{eqnarray}\label{3.2}
g(k)=\frac{4\pi e^2}{\epsilon k}\left(1-e^{-kd}\right).
\end{eqnarray}
Here $\epsilon$ is the background dielectric constant of the medium. We do not distinguish the QWs dielectric constants and a barrier between them for simplicity. If $kd<<1$, the exciton-exciton interaction can be considered as a contact one, $g\approx 4\pi e^2d/\epsilon$. This form of the contact exciton interaction potential is assumed in eq.(\ref{3.1}) and everywhere below. Taking into account eq.(\ref{9}) and eq.(\ref{15}) the screened potential is $W(k)=U(k)/\epsilon(k)$. Here the dielectric permittivity reads
\begin{eqnarray}\label{3.3}
\frac{1}{\epsilon(k)}=1+gP^c_{k,0}+g\frac{P^n_{k,0}[1+gP^c_{k,0}]}{1-3gP^n_{k,0}}=\left(1+gP^c_{k,0}\right)\frac{1-2gP^n_{k,0}}{1-3gP^n_{k,0}}.
\end{eqnarray}
The spatial dependence of the screened potential yields
\begin{eqnarray}\label{3.4}
W(\textbf{r})=\int \frac{d\textbf{k}}{(2\pi)^2}e^{i\textbf{kr}}\frac{U(k)}{\epsilon(k)}=\int_0^{\infty}\frac{kdk}{2\pi}J_0(kr)\frac{U(k)}{\epsilon(k)}.
\end{eqnarray}
Let us consider the impurity located at distance $H$ from the excitonic layer. We have
\begin{eqnarray}\label{3.5}
U(k)=\frac{2\pi e^2}{\epsilon k}\left(1-e^{-kd}\right)e^{-kH}\approx\frac{2\pi e^2 d}{\epsilon}e^{-kH}.
\end{eqnarray}

\begin{figure} [t]
\centerline{\input epsf \epsfysize=3.5cm \epsfbox{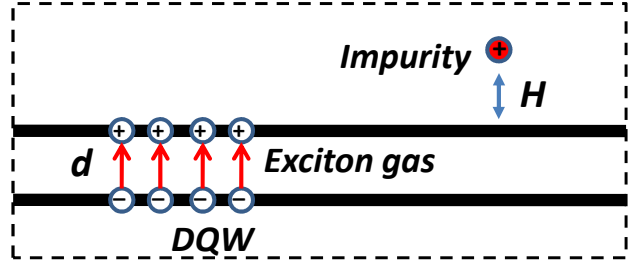}}
\caption{Sketch of the system under study.}
\label{Figure1}
\end{figure}

Here the last equality is taken with the same accuracy $(kd<<1)$ as the exciton-exciton interaction potential (see text after eq.(\ref{3.2})). We consider the screened potential at the long-wavelength limit, $r>>s/T$. This inequality is an equivalent of $k<<p\sim T/s$ in eq.(\ref{19}). For small $k$, we find from eq.(\ref{19})
\begin{eqnarray}\label{3.6}
P^{n(1)}_{k,0}=\frac{g^2n_c^2}{2\pi s^2}\frac{T}{s^2k^2}
\end{eqnarray}
Taking into account that
\begin{eqnarray}\label{3.7}
P^c_{k,0}=-\frac{2n_c}{2gn_c+k^2/2M};\,\,\,\,\,
P^{n(0)}_{k,0}=-\frac{g^2n_c^2}{4s^2}\frac{1}{sk},
\end{eqnarray}
the asymptotic form of the screened potential can be easily evaluated from eq.(\ref{3.4}). Expending $\epsilon^{-1}(k)$ in powers of $k$, and calculating integrals, we come to the result
\begin{eqnarray}\label{3.8}
W(r)=\frac{2}{3}\frac{e^2d}{4\epsilon Mgn_c}\frac{3H(2H-3r^2)}{(H^2+r^2)^{7/2}}-\\\nonumber
-\frac{e^2d}{4\epsilon Mgn_c}\left(\frac{2}{3}\frac{1}{4Mgn_c}+\frac{2\pi s^4}{9g^3n_c^2T}\right)
\frac{15H(8H^4-40r^2H^2+15r^4)}{(H^2+r^2)^{11/2}}-\\\nonumber
-\frac{e^2d}{144\epsilon}\frac{(2\pi)^2s^5}{g^4n_c^3T^2}\frac{45(16H^6-120H^4r^2+90H^2r^4-5r^6)}{(H^2+r^2)^{13/2}}.
\end{eqnarray}
If $H<<r$ we can insert $H=0$ into eq.(\ref{3.8}) and one finds the simplified result
\begin{eqnarray}\label{3.9}
W(r)=\frac{225e^2d}{144\epsilon}\frac{(2\pi)^2s^5}{g^4n_c^3T^2}\frac{1}{r^7}.
\end{eqnarray}
It should be noted, that we consider here the exciton BEC with point-like inter-particle interaction $g=const$, in contrast to our previous work \cite{KovalevChaplik1}, where we consider the general case $g(\textbf{r})$ at zero temperatures. It was shown there, that, at $H=0$ the screened potential at the distances $r^2>>a^\ast/16\pi n_cd$, where $a^\ast=\epsilon/Me^2$, also has $1/r^7$ damping law with the temperature-independent coefficient. At finite $T$, as it can be seen from eq.(\ref{3.9}), the 'tail' of the screened potential at large distances $r>>s/T$ is substantially determined by the temperature.

\begin{figure} [t]
\centerline{\input epsf \epsfysize=6cm \epsfbox{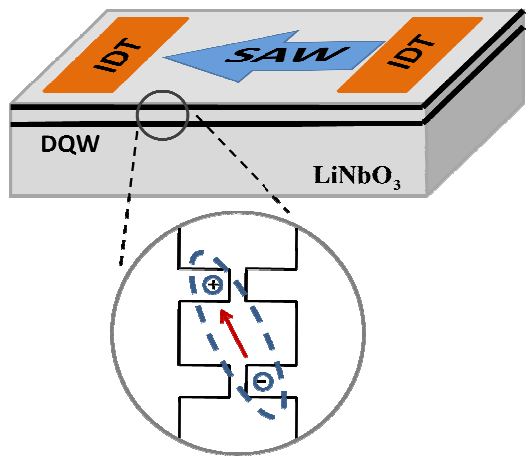}}
\caption{Schematic view of the experimental setup.
Interdigital transducer (IDT) produces the SAW traveling on the surface of the crystal. Electric field created by SAW due to piezoelectric effect interacts with dipole moments of the excitons.}
\label{Figure2}
\end{figure}

\section{Surface acoustic wave absorption}

We consider a semi-infinite $LiNbO_3$ piezoelectric crystal as a substrate having a point-group symmetry $C_{6v}$. DQW is on top of this substrate crystal, see Fig. 2. Surface $(x,z)$ contains the symmetry axis of the substrate crystal. The surface acoustic wave propagates along the $x$-axis orthogonal to the symmetry axis oriented along $z$ direction (Blushtain-Gulyaev wave). In our picture, the displacement vector has the components $\textbf{u}=(0,0,u(x,y))$. The scalar potential contains two contributions due to both the piezoelectric crystal and the total exciton density $\delta N_{k\omega}$ variation.  The dependence of $u$ and $\varphi$ on time and positions are given by the system of equations in the medium ($\varphi=\varphi^{(i)}$, for $y>0$)
\begin{eqnarray}\label{4.1}
\rho u^{''}_{tt}=\lambda\Delta u-\beta\Delta \varphi^{(i)},\\\nonumber
\varepsilon\Delta\varphi^{(i)}+4\pi\beta\Delta u=0,
\end{eqnarray}
and outside the medium ($\varphi=\varphi^{(e)}$, for $y<0$)
\begin{eqnarray}\label{4.2}
\Delta\varphi^{(e)}=0.
\end{eqnarray}
Here $\rho,\lambda,\beta,\varepsilon$ are density, elastic modulus, piezoelectric constant and dielectric constant of the substrate crystal, respectively. Dynamic equations (\ref{4.1}) and (\ref{4.2}) have to be supplemented with the boundary condition which is expressed by $\sigma_{zy}|_{y=0}=0$, where $\sigma_{zy}$ is a stress tensor. Moreover, the Poisson equation in eq.(\ref{4.1}) needs boundary conditions for electric induction vector $\textbf{D}$ and electric potential $\varphi$. From the electrostatic point of view, the exciton layer can be viewed as an electric double layer. To apply this model, conditions $kd<<1$ and $k\varrho<<1$ must be satisfied. Here $\varrho$ is a damping decrement of the displacement vector in the $y$-direction. The boundary conditions for the double electric layer at point $y=0$ have the form
\begin{eqnarray}\label{4.3}
D_y^{(i)}=-\frac{\partial\varphi^{(i)}}{\partial y};\,\,\,\varphi^{(e)}-\varphi^{(i)}=4\pi p\delta N,
\end{eqnarray}
where $p=ed$ is an absolute exciton dipole moment value. Taking into account all this boundary conditions, we come to the following dispersion relation
\begin{eqnarray}\label{4.4}
1-\frac{\gamma k}{(\epsilon+1)\sqrt{k^2-\omega^2/c^2}}+b\gamma k P_{k,\omega}=0.
\end{eqnarray}
Here $b=4\pi e^2d^2/(\epsilon+1)$ and $\gamma=4\pi\beta^2/\epsilon\rho c^2$ -- piezoelectric coupling coefficient. This dispersion equation eq.(\ref{4.4}) can be rewritten in a clearer form
\begin{eqnarray}\label{4.5}
\sqrt{k^2-\omega^2/c^2}=\frac{\gamma k}{(\epsilon+1)}\frac{1}{(1+b\gamma k P_{k,\omega})}.
\end{eqnarray}
SAW damping is given by the imaginary part of the wave vector $k(\omega)=k_1+ik_2$ in eq.(\ref{4.5}). The general analytical solution for $k(\omega)$ cannot be obviously found. To find the SAW damping, we apply the method of successive approximations. In the absence of excitons $P_{k,\omega}=0$ the solution of eq.(\ref{4.5}) has only the real part $k_1=\omega/\tilde{c}$, where $\tilde{c}=c\sqrt{1-\gamma^2/(\epsilon+1)^2}$ is a speed of SAW renormalized due to the piezoelectric effect. In the presence of excitons we assume that $k_2<<k_1$ and
\begin{eqnarray}\label{4.6}
k_2=-\frac{b\gamma^3k_1^2}{(\epsilon+1)^2}\frac{P_2}{(1+b\gamma k_1 P_1)^2+(b\gamma k_1 P_2)^2}
\end{eqnarray}
where $P_{k,\omega}=P_1+iP_2$ and $k=k_1=\omega/\tilde{c}$ has to be inserted in the r.h.s. of eq.(\ref{4.6}). As we see, the SAW damping coefficient is determined by the imaginary part of the exciton BEC polarization operator. In the lowest order on a weak exciton-exciton interaction constant $g$, it is given by
\begin{eqnarray}\label{4.8}
P_{k,\omega}=P^c_{k\omega}+P^n_{k\omega}\frac{1+gP^c_{k\omega}}{1-3gP^n_{k\omega}}\approx P^c_{k\omega}+P^n_{k\omega}.
\end{eqnarray}
The imaginary part of $P^c_{k\omega}$ has the form (see eq.(\ref{9})) $Im \,P^c_{k\omega}\sim\delta(\omega^2-s^2k^2)$ and is only relevant for $\omega=\tilde{c}k=sk$. If $\tilde{c}\neq s$, the SAW damping is given by the second term in eq.(\ref{4.8}), which has temperature-independent $P^{n(0)}_{k,\omega}$ and temperature-dependent $P^{n(1)}_{k,\omega}$ contributions (see eq.(\ref{19}) and eq.(\ref{20}) above). The real and imaginary parts of the first, temperature-independent $P^{n(0)}_{k,\omega}$ term, are given by eq.(\ref{23}). The imaginary part of the second temperature-dependent $P^{n(1)}_{k,\omega}$ term can be found from eq.(\ref{22}). For the quantum regime $sk>>T$, simple, but lengthy calculations yield
\begin{eqnarray}\label{4.9}
Im\,P^{n(1)}_{k,\omega}=-\frac{4\pi g^2n_c^2}{s^2}\frac{T}{sk}\frac{\theta(\omega^2-s^2k^2)}{\sqrt{\omega^2-s^2k^2}}e^{-\frac{|\omega|-sk}{2T}}-\\\nonumber
-\frac{2\pi g^2n_c^2}{s^2}\sqrt{\frac{2\pi T}{sk}}\frac{\theta(s^2k^2-\omega^2)}{\sqrt{s^2k^2-\omega^2}}\left(e^{-\frac{sk-|\omega|}{2T}}-e^{-\frac{|\omega|+sk}{2T}}\right)
\end{eqnarray}
The real part of $P^{n(1)}_{k,\omega}$ is small (due to factor $sk>>T$) in comparison with the real parts of both $P^{n(0)}_{k,\omega}$ and $P^c_{k\omega}$ and we ignore it for simplicity. Let us present here the total expression for $P_{k,\omega}$. If $\omega^2>s^2k^2$, we have
\begin{eqnarray}\label{4.10}
\hspace{-1cm}P_{k,\omega}=\frac{n_ck^2/M}{\omega^2-s^2k^2}-i\frac{g^2n_c^2}{4s^2}\frac{\theta[\omega^2-s^2k^2]}{\sqrt{\omega^2-s^2k^2}}
-i\frac{4\pi g^2n_c^2}{s^2}\frac{T}{sk}\frac{\theta[\omega^2-s^2k^2]}{\sqrt{\omega^2-s^2k^2}}e^{-\frac{|\omega|-sk}{2T}}
\end{eqnarray}
and for $\omega^2<s^2k^2$
\begin{eqnarray}\label{4.11}
\hspace{-0cm}P_{k,\omega}=-\frac{n_ck^2/M}{s^2k^2-\omega^2}-\frac{g^2n_c^2}{4s^2}\frac{\theta[s^2k^2-\omega^2]}{\sqrt{s^2k^2-\omega^2}}\\\nonumber
-i\frac{2\pi g^2n_c^2}{s^2}\sqrt{\frac{2\pi T}{sk}}\frac{\theta[s^2k^2-\omega^2]}{\sqrt{s^2k^2-\omega^2}}\left(e^{-\frac{sk-|\omega|}{2T}}-e^{-\frac{|\omega|+sk}{2T}}\right).
\end{eqnarray}
To find the SAW damping, one needs to insert these expressions (taken at $k=k_1$) in eq.(\ref{4.6}). The result is very cumbersome, and we present here the limiting cases. If $|\tilde{c}^2-s^2|>>b\gamma n_c\omega/M\tilde{c}$ one has $b\gamma k_1P_{1,2}<<1$. Inserting eq.(\ref{4.10}) and eq.(\ref{4.11}) in eq.(\ref{4.6}), we find the SAW damping coefficient
\begin{eqnarray}\label{4.12}
\hspace{-1cm}k_2=\frac{b\gamma^3\omega}{\tilde{c}^2(\epsilon+1)^2}\frac{g^2n_c^2}{4s^2}\frac{1}{\sqrt{1-s^2/\tilde{c}^2}}
\left(1+\frac{16\pi T \tilde{c}}{\omega s}e^{-\frac{\omega(\tilde{c}-s)}{2T\tilde{c}}}\right),\,\,\,\,\,\,\tilde{c}>s;\\\nonumber
\hspace{-1cm}k_2=\frac{b\gamma^3\omega}{\tilde{c}^2(\epsilon+1)^2}\frac{2\pi g^2n_c^2}{s^2}\frac{1}{\sqrt{s^2/\tilde{c}^2-1}}
\sqrt{\frac{2\pi T \tilde{c}}{\omega s}}\left(e^{-\frac{\omega(s-\tilde{c})}{2T\tilde{c}}}-e^{-\frac{\omega(s+\tilde{c})}{2T\tilde{c}}}\right),\,\,\,\tilde{c}<s.
\end{eqnarray}
In the opposite limit $|\tilde{c}^2-s^2|<<b\gamma n_c\omega/M\tilde{c}$, one has $b\gamma k_1P_{1}>>1$ and $P_{1}>>P_{2}$. The SAW damping reads
\begin{eqnarray}\label{4.13}
\hspace{-1cm}k_2=\frac{\gamma(gM\tilde{c})^2\omega}{4s^2b(\epsilon+1)^2}(1-s^2/\tilde{c}^2)^{3/2}\left(1+\frac{16\pi T \tilde{c}}{\omega s}e^{-\frac{\omega(\tilde{c}-s)}{2T\tilde{c}}}\right),\,\,\,\,\,\,\tilde{c}>s;\\\nonumber
\hspace{-1cm}k_2=\frac{2\pi\gamma(gM\tilde{c})^2\omega}{s^2b(\epsilon+1)^2}(s^2/\tilde{c}^2-1)^{3/2}\sqrt{\frac{2\pi T \tilde{c}}{\omega s}}\left(e^{-\frac{\omega(s-\tilde{c})}{2T\tilde{c}}}-e^{-\frac{\omega(s+\tilde{c})}{2T\tilde{c}}}\right),\,\,\,\tilde{c}<s.
\end{eqnarray}
We see that in both cases at $T=0$ $k_2\sim\omega\theta(\tilde{c}^2-s^2)$ and $k_2=0$ if $s^2>\tilde{c}^2$, in other words, $k_2$ grows up linearly with increasing the SAW frequency and has specific singularity versus exciton density. Indeed, SAW damping occurs for $\tilde{c}^2>s^2$ while $s^2=gn_c/M$. Thus, inequality $\tilde{c}^2>s^2$  is equivalent to  $n_c<n_c^0$, where critical exciton density is given by $n_c^0=M\tilde{c}^2/g$.

We conclude from the form of damping eq.(\ref{4.12}) and eq.(\ref{4.13}) that, for $T\neq 0$, the SAW damping occurs also for $s>\tilde{c}$ due to the Landau damping processes of the noncondensate particles. But, nevertheless, both temperature-dependent terms in eq.(\ref{4.12}) and eq.(\ref{4.13}) have an exponentially small contribution at sufficiently low T. In equation (\ref{4.13}) it is also shown that the SAW damping (at $T=0$) goes to zero when $n_c\rightarrow n_c^0$ as $k_2\sim\theta(n_c^0-n_c)(n_c^0-n_c)^{3/2}$.

Eqs.(\ref{4.12}) and (\ref{4.13}) are our main results of this section.

\section{Conclusion}
We have developed the theory of the excitonic BEC response to the static and dynamical perturbations. This response theory has been applied to studying the screening phenomena and SAW damping in the excitonic BEC. We demonstrate for the first time that the spatial behavior of the impurity screened potential substantially differs from the one known for a 2D electron gas system. The reason of this phenomena is the bosonic nature of the excitons and the possibility of their condensation.

The SAW damping, as it was shown here, has specific and unusual behavior vs excitonic density. It should be stressed out that such a behavior is held only in the presence of the excitonic BEC. Indeed, in the absence of BEC, exciton gas absorbs the SAW and its damping is a monotonic function of the exciton density \cite{KovalevChaplik}. We assume that such a threshold-like effect can be used not only for excitonic BEC experimental detection, but also for studying BEC of other bose-type particles, such as exciton-polaritons, for example.

Our research work has been supported by the RFBR Foundation (projects 12-02-31012 and 11-02-00060).

\end{document}